\documentclass[runningheads]{llncs}
\usepackage{graphicx}
\usepackage[ruled,linesnumbered]{algorithm2e}
\usepackage{float}
\usepackage{csquotes}
\usepackage[table]{xcolor}

\newcommand{\specialcell}[2][c]{%
  \begin{tabular}[#1]{@{}c@{}}#2\end{tabular}}

\definecolor{Gray}{gray}{0.9}

\begin{document}
\title{Test Case Prioritization Using Test Similarities}
%
%

\author{
Alireza Haghighatkhah \orcidID{0000-0002-1085-5825}\and Mika M\"{a}ntyl\"{a}\orcidID{0000-0002-2841-5879}\and Markku Oivo \orcidID{0000-0002-1698-2323}\and Pasi Kuvaja \orcidID{0000-0002-1488-6928}
}

%
%
\institute{M3S Research Unit, University of Oulu, P.O. Box 3000, 90014 Oulu, Finland.
\email{\{firstname.lastname\}@oulu.fi}}
\maketitle              
\begin{abstract}
A classical heuristic in software testing is to reward diversity, which implies that a higher priority must be assigned to test cases that differ the most from those already prioritized. This approach is commonly known as similarity-based test prioritization (SBTP) and can be realized using a variety of techniques. The objective of our study is to investigate whether SBTP is more effective at finding defects than random permutation, as well as determine which SBTP implementations lead to better results. To achieve our objective, we implemented five different techniques from the literature and conducted an experiment using the defects4j dataset, which contains 395 real faults from six real-world open-source Java programs. Findings indicate that running the most dissimilar test cases early in the process is largely more effective than random permutation (Vargha\textendash Delaney A [VDA]: 0.76\textendash0.99 observed using normalized compression distance). No technique was found to be superior with respect to the effectiveness. Locality-sensitive hashing was, to a small extent, less effective than other SBTP techniques (VDA: 0.38 observed in comparison to normalized compression distance), but its speed largely outperformed the other techniques (i.e., it was approximately 5\textendash111 times faster). Our results bring to mind the well-known adage, \enquote{don't put all your eggs in one basket}. To effectively consume a limited testing budget, one should spread it evenly across different parts of the system by running the most dissimilar test cases early in the testing process.

\end{abstract}
\section{Introduction}

The software industry is moving toward an agile, continuous delivery paradigm in which software changes are released more frequently and considerably faster than before \cite{rodriguez2017continuous}. This development paradigm has brought many benefits but posed several challenges, particularly regarding software quality \cite{rodriguez2017continuous,mantyla2015rapid}. To ensure software correctness, software developers employ regression testing (RT), which involves running a dedicated regression test suite after each revision to verify that recent changes have not negatively impacted the software's functionality \cite{engstrom2010qualitative}. Industrial software-intensive systems often comprise many test cases, and the execution of these test cases require several hours or even days. For example, the JOnAS Java EE middleware requires running 43,024 test cases to verify all of its 16 configurations \cite{kessis2005experiences}. To improve RT processes, the software engineering literature has proposed many solutions \cite{yoo2012regression}. Test case prioritization (TCP) \cite{rothermel2001prioritizing} is one of these solutions; it is concerned with the ideal ordering of test cases to maximize desirable properties (i.e., early fault detection). From the fault detection viewpoint, TCP seems to be a safe approach because it does not eliminate test cases and simply permutes them within the test suite.

To increase the likelihood of detecting faults, one potential strategy is spreading the testing budget evenly across different parts of the system \cite{jiang2009adaptive,hemmati2010reducing,ledru2012prioritizing}, and realizing this strategy involves utilizing a diverse set of test cases. To devise a diverse test suite, one needs to measure similarities among the test cases. The notion of similarity measurement is a subject of interest for many applications. The degree to which two objects share characteristics is called similarity, and the degree to which they differ is termed distance. In the software testing literature, a point of particular interest is quantifying similarities among test cases. For example, in coverage-based testing, coverage information has been used as a proxy to measure the similarities among test cases \cite{jiang2009adaptive}. More recently, several other properties have been described in the literature i.e., the overlap between test paths and their coverage in model-based testing \cite{hemmati2011empirical,cartaxo2011use}, as well as the source code of test cases \cite{ledru2012prioritizing}, test input and output \cite{henard2016comparing}, topic models extracted from test scripts \cite{thomas2014static}, and even English document of manual test cases \cite{hemmati2017prioritizing}.

The main intuition is that \textit{test cases that capture the same faults tend to be more similar to each other, and test cases that capture different faults tend to be more dissimilar} \cite{jiang2009adaptive,hemmati2010reducing,ledru2012prioritizing}. The number of published empirical studies that support this intuition are growing (e.g., \cite{hemmati2013achieving,hemmati2017prioritizing,arafeen2013test,thomas2014static,feldt2016test,flemstrom2017similarity}). The implication for TCP is that a higher priority must be assigned to test cases that are most dissimilar to those already prioritized. This can be realized by maximizing the distances among test cases ordered in the test suite. Similarity-based test prioritization (SBTP) is a black-box static technique (i.e., it does not require the source code and execution of the system under test) that can potentially be applied, for example, where code instrumentation is too costly or impossible.

The natural question that arises is whether running the most dissimilar test cases early in the testing process improves the test suite's fault-detection capability. SBTP can be implemented in a variety of ways, such as applying different similarity metrics. Thus, a follow-up question that arises is which implementation yields the best results. A similar objective was pursued by Ledru et al. \cite{ledru2012prioritizing} in 2012. The authors conducted a comprehensive experiment on the Siemens test suite and evaluated four classical string metrics using a pairwise algorithm. This study extends prior research by investigating the effectiveness and performance of five different SBTP techniques (4 additional in comparison to Ledru et al.). These techniques rely on different similarity metrics and were selected from the literature based on the results of recent experimental studies \cite{feldt2008searching,hemmati2010industrial,feldt2016test,ledru2012prioritizing,miranda2018fast}.

The ultimate objective of our study is to detect regression faults early in the testing process, allowing software developers to perform RT more frequently and continuously. To achieve this objective, we conducted an experiment using the defects4j dataset \cite{just2014defects4j}, which contains 395 real faults from 6 real-world open-source Java programs. Findings from our study indicate that:

\begin{itemize}
\item When their average percentage of faults detected (APFD) are compared, test suites ordered by SBTP are largely more effective than random permutation (VDA: 0.76\textendash 0.99 observed using normalized compression distance [NCD] across all subjects), which means running the most dissimilar test cases early in the testing process improves the test suite's fault-detection capability.
\item Of the 5 SBTP implementations investigated, no technique was found to be superior with respect to the effectiveness. Locality-sensitive hashing (LSH) was, to a limited extent, less effective than other SBTP techniques (VDA: 0.38 observed in comparison to NCD), but its speed largely outperformed the other techniques (i.e., it was approximately 5\textendash111 times faster).

\end{itemize}

Our findings yield important academic and practical implications. From the academic perspective, we provide empirical evidence that supports test diversity and its impact on TCP. From the of practitioners’ perspective, our results bring to mind the well-known adage, \enquote{don't put all your eggs in one basket}. To effectively consume a limited testing budget, one should spread it evenly across different parts of the system by running the most dissimilar test cases early in the process. The remainder of the paper is organized as follows. Section 2 discusses the background and related works. Section 3 describes the research methodology, and Section 4 presents answers to the research questions. The findings are discussed in Section 5, and conclusions are discussed in Section 6.

\section{Background and Related Work}
\subsection{Background}
\begin{figure}[h]
\centering
\includegraphics[width=\textwidth]{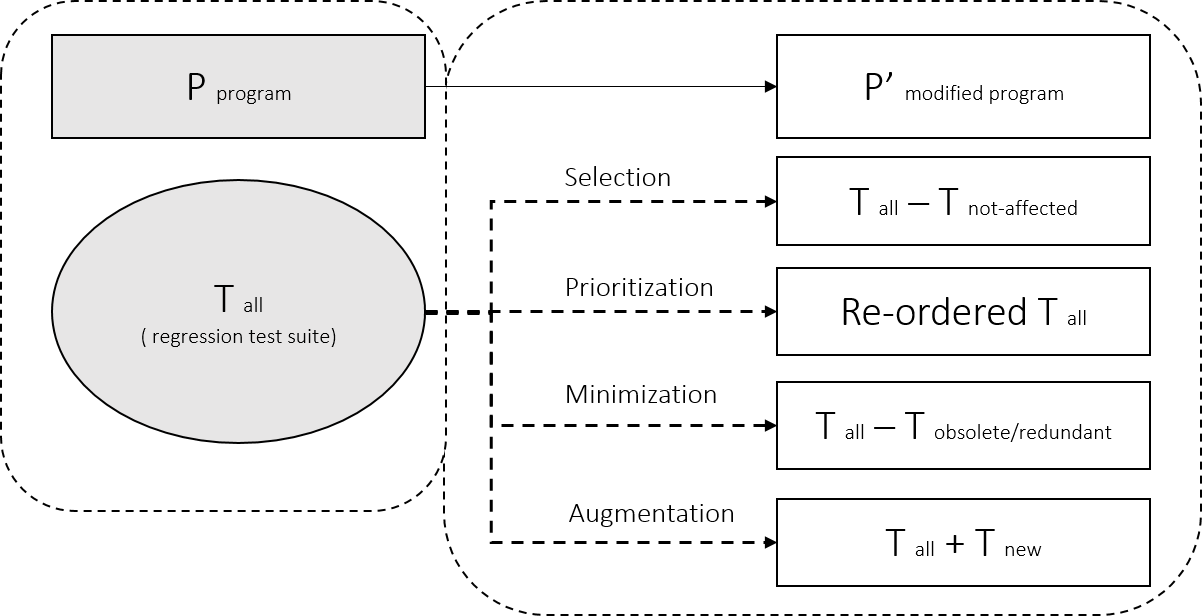}
\caption{General model of RT techniques} \label{fig:RT_Background}
\end{figure}

Figure \ref{fig:RT_Background} presents a general model of RT techniques. Let $P$ be a program, $P'$ be a modified version of the program, and $T$ be a test suite developed for $P$. In the transition from $P$ to $P'$, a previously verified behavior of $P$ may have become faulty in $P'$. RT seeks to capture regressions in $P'$ and verify that changes to the system have not negatively impacted any previously verified functionalities. During RT, several techniques may be employed. One of the techniques is test suite minimization; it seeks to identify and permanently eliminate obsolete or redundant test cases from the test suite. Another technique, regression test selection, aims to select only the subset of test cases affected by the recent changes. TCP is concerned with the ideal ordering of test cases to maximize desirable properties (i.e., early fault detection), while test suite augmentation aims to identify newly added source code and generate new test cases accordingly.

\subsection{Related Work}

A similarity metric, which is also known as similarity/distance function, is a metric that measures the similarity or distance (i.e., inverse similarity) between two objects. Similarity metrics have been widely applied in the literature (e.g., classification problems, plagiarism detection, sequence and image analysis). 

In software engineering, particularly in software testing, similarity metrics have been applied. For instance, Shahbazi and Miller \cite{shahbazi2016black} conducted a large empirical study on black-box automated test-case generation using several string metrics. Their results indicate that superior test cases can be generated by controlling the diversity and length distribution of the string test cases. Hemmati et al. \cite{hemmati2013achieving} proposed a similarity-based test case selection technique that selects the most diverse subset of test cases among those generated by applying a coverage criterion on a test model. Feldt et al. \cite{feldt2016test} proposed the test set diameter (TSDm) technique, which was developed based on NCD for multisets. Their results indicate that test selection using TSDm leads to higher structural and fault coverage than random selection. NCD multisets, which provides similarity measurement at the level of entire sets of elements rather than between pairs, have also been applied in the TCP literature recently \cite{henard2016comparing}.

To implement SBTP, the distances among test cases must be measured using a specific metric, and this information must then be leveraged to perform TCP. Ledru et al. \cite{ledru2012prioritizing} conducted a comprehensive experiment on the Siemens test suite and evaluated four classical string metrics for TCP purposes (i.e., Cartesian, Levenshtein, Hamming, and Manhattan distance). Their findings indicated that TCP using string metrics is more effective than random prioritization, and on average, Manhattan distance yields better results than the other investigated metrics. To calculate the distance between a test case $t$ and set of test cases $T'$, Ledru et al. proposed the following function, which uses distance measure $d$: 

$$distance(t, T', d) = min \{d(t, t_i)| t_i \in T', t_i \neq t\} $$

Ledru et al. used the min operation because an empirical study by Jiang et al. \cite{jiang2009adaptive} showed that maximize-minimum is more efficient than maximize-average and maximize-maximum. Ledru et al. also proposed an algorithm (algorithm \ref{label_pairwise_algorithm}) that iteratively picks the most dissimilar test case (i.e., having the greatest distance from a set of already prioritized test cases).

\begin{algorithm}[h]
\label{label_pairwise_algorithm}
\SetAlgoLined
\KwData{Test Suite $TS$}
\KwResult{Prioritized Suite $PS$}
Find $t\in TS$ with the maximum $distance(t, TS)$\;
Append $t$ to $PS$ and remove from $TS$\;
 \While{$TS$ is not empty}{
  Find $t\in TS$ with the maximum $distance(t, PS)$\;
  Append $t$ to $PS$ and remove from $TS$\;
 }
 \caption{Similarity-based TCP Using a Pairwise Algorithm}
\end{algorithm}

Using SBTP with a pairwise algorithm comes with the cost of pairwise comparison, and its performance becomes inefficient as the test suite becomes larger. The underlying issue in SBTP can be defined as a similarity search problem, which involves searching within a large set of objects for a subset of objects that closely resemble a given query object. One popular approach to solving similarity search problems is LSH, which was originally introduced by Indyk and Motwani \cite{indyk1998approximate} in 1998. LSH hashes input items so that similar items map to the same buckets with high probability \cite{leskovec2014mining}. LSH is widely used in the literature (see the many references in Google Scholar to \cite{indyk1998approximate}) but is only occasionally applied to software engineering problems (e.g., clone detection \cite{jiang2007deckard} and test generation \cite{shahbazi2016black}). More recently, Miranda et al. \cite{miranda2018fast} proposed an approach based on LSH, which provides scalable SBTP in both white-box and black-box fashion.

The purpose of our study is to investigate whether SBTP is more effective at finding defects than random permutation and which SBTP implementations yield the best results. A similar objective was pursued by Ledru et al. \cite{ledru2012prioritizing} in 2012. In comparison to their work, we have investigated five different techniques with respect to their effectiveness and performance. These techniques rely on different similarity metrics and were selected from the literature based on the results of recent experimental studies \cite{feldt2008searching,hemmati2010industrial,feldt2016test,ledru2012prioritizing,miranda2018fast}. The rationale behind their selection and details about their implementation is described in section \ref{label_studydesign}. The Siemens test suite, which was used by Ledru et al., is a classical dataset and widely used in the software testing literature. However, its representative character has been debated for several reasons (e.g., in \cite{orso2014software}, which was also acknowledged by Ledru et al. in \cite{ledru2012prioritizing}). In this work, we report an experiment conducted on the defects4j dataset \cite{just2014defects4j}, which contains 395 real faults from 6 open-source Java programs.

\section{Research Method}
In this section, the study's objective and research questions, study subject, study design, and evaluation methods are discussed.

\subsection{Objective and Research Questions}
The main objective of our study is to catch regression faults early in the testing process, allowing software developers to perform RT more frequently and continuously. The research questions and their rationales are as follows:

\textbf{RQ1: Is prioritization by similarity-based TCP more effective at finding defects than random permutation?} This research question is designed to investigate whether running the most dissimilar test cases early in the testing process improves the test suite's fault-detection capability in comparison to random ordering.

\textbf{RQ2: Which similarity-based TCP technique is the most effective and has the best performance?} This research question is designed to compare the effectiveness and performance of investigated SBTP implementations. The rationale behind the investigated techniques' selection and details about their implementation are described in section \ref{label_studydesign}.

\subsection{Subjects under study}
To answer our research questions, we conducted an experiment using the defects4j dataset \cite{just2014defects4j}, which contains 395 real faults from 6 real-world open-source Java programs. The subject's characteristics are presented in Table \ref{table:subjects}. Each analyzed subject's name is presented in the first column, while the second column shows the number of versions analyzed for each program. The third and fourth columns present the median number of test classes and test cases, and the range is in parentheses. The last two columns show the source's size (kilo line-of-code) and test code for the most recent version, as reported by SLOCCount \footnote{SLOCCount is a suite of programs used to count lines of code: \url{https://www.dwheeler.com/sloccount/}}.

\begin{table}[h]
\centering
\caption {Subject Characteristics}
\label{table:subjects}
\begin{tabular}{|l|l|l|l|l|l|}
\hline                       
 Project & Versions & Test classes & Test cases & S-LOC & T-LOC \\
\hline
\hline 
JFreeChart (Chart) & 26 & 323 (301\textendash356) & 1789 (1591\textendash2193) & 123.527 & 37.396 \\ 
Closure Compiler (Closure) & 133 & 216 (118\textendash235) & 7389 (2595\textendash8443) & 251.855 & 85.138 \\
Apache Lang (Lang) & 65 & 89 (81\textendash111) & 1760 (1540\textendash2291) & 45.609 & 28.199 \\
Apache Math (Math) & 106 & 253 (91\textendash385) & 2319 (817\textendash4378) & 22.738 & 12.238 \\
Mockito & 38 & 237 (128\textendash268) & 1233 (704\textendash1388) & 38.914 & 10.638 \\
Joda-Time (Time) & 27 & 122 (120\textendash123) & 3906 (3749\textendash4041) & 176.965 & 41.536 \\
\hline
\end{tabular}
\end{table}

\subsection{Study design}
\label{label_studydesign}

To answer RQ1, we compared the effectiveness of SBTP with random permutation. SBTP does not use a system under test; thus, it can hardly be more effective than TCP techniques, which use code coverage criteria \cite{ledru2012prioritizing}. Thus, like Ledru et al., we used random permutation as the baseline of our experiment. For the sake of a sanity check, we also included a TCP approach in which we minimize the diversity (i.e., maximize similarity among test cases). The rationale behind our sanity check is if diversity is valuable in TCP, then minimizing diversity should, in turn, negatively affect the test suite's fault-detection capability \cite{henard2016comparing}. Effectiveness was measured using APFD, which is a commonly used metric in the TCP literature and elaborated on in section \ref{label_evaluation}. 

To answer RQ2, we presented the aggregated the investigated techniques' performance and effectiveness within and across studied subjects. Using the aggregated values, we can determine which technique achieved the best effectiveness and performance on average. The five SBTP techniques presented in Table \ref{table:investigated_techniques} were selected from the literature and investigated in this experiment. To calculate the distances, we automatically downloaded the source code for all studied versions and used the source code behind the test classes at their exact version.
 
\begin{table}[h]
\centering
\caption {TCP Techniques Investigated}
\label{table:investigated_techniques}
\begin{tabular}{|l|l|l|l|}
\hline                       
 Name (Acronym) & Objective & Reference \\
\hline
\hline 
Random Permutation (RND) & Baseline & - \\ 
Manhattan Distance (MNH) & Maximize diversity & Ledru et al. \cite{ledru2012prioritizing} \\ 
Jaccard Distance (JAC) & Maximize diversity & Hemmati \& Briand \cite{hemmati2010industrial} \\ 
Normalized Compression Distance (NCD) & Maximize diversity & Feldt et al. \cite{feldt2008searching} \\ 
Sanity Check (SC) using NCD & Maximize similarity & - \\ 
NCD Multisets (NCD-MS) & Maximize diversity & Feldt et al. \cite{feldt2016test} \\ 
Locality Sensitive Hasing (LSH) & Maximize diversity & Miranda et al.\cite{miranda2018fast} \\ 
\hline
\end{tabular}
\end{table} 
 
We implemented the Manhattan, Jaccard , NCD, and NCD Multisets using the pairwise algorithm proposed by Ledru et al. \cite{ledru2012prioritizing}. The Manhattan distance between two objects is the sum of the differences of their corresponding components. To calculate the Manhattan distance, the source code is converted to a vector of numbers. In practice, each character should be replaced with their ASCII code (or any other numerical coding). The Jaccard similarity between two sets $x$ and $y$ is defined as $JS(x,y)= |x \cap y|/|x \cup y|$, and their distance is $JD(x, y) = 1- JS(x, y)$. To calculate the Jaccard distance, the source code is converted to a set of k-shingles (e.g., any substring of length k found within the text). In our study, we used $k=5$, which is commonly used in the analysis of relatively short documents \cite{leskovec2014mining}.

NCD and NCD Multisets both rely on a compressor function $C$, which calculates the approximate Kolmogorov complexity and returns the length of the input string after its compression, using a chosen compression program. In this study, we used LZ4, which is a high-speed lossless data compression algorithm \footnote{The LZ4 compression algorithm and details regarding its implementation are available at \url{http://lz4.github.io/lz4/}}. The difference between NCD and NCD Multisets is that the latter performs similarity measurement at the level of entire sets of elements rather than between pairs. For the NCD Multisets, we adapted the pairwise algorithm so that at each iteration, we pick a test $t\in TS$ that has maximum Kolmogorov complexity when compressed with the entire set of the already prioritized suite $PS$. This means that the candidate test has less mutual information with $PS$ and is more different than any other $t\in TS$.

Furthermore, we implemented LSH using the MinHash technique to rapidly estimate Jaccard similarity. In our implementation, we followed the instructions provided by \cite{leskovec2014mining}, which are also described here. To estimate the Jaccard similarity, we converted the source code to a set of 5-shingles. However, their size is often large, and it is impractical to use them directly. Using MinHashing technique, we replaced these sets with a much smaller representation (e.g., a signature) while preserving the Jaccard similarity between them. Given a hash function $h$ and an input set $S$, we hashed all elements in the set using the hash function and picked the minimum resulting value as MinHash of $S$. This process was repeated $P$ times (i.e., the number of permutations) using different hash functions to calculate the signature of a set (e.g., a sequence of MinHashes). Thereafter, the Jaccard similarity of two sets can be estimated using the fraction of common MinHashes in their signature. Using MinHashing, we were able to compress large sets into a small signature; similarity searches among large numbers of pairs is inefficient.

LSH works with a signature matrix (e.g., MinHash signatures as column) and divides it into $b$ bands consisting of $r$ rows each. For each band, LSH takes vectors of numbers (e.g., the portion of one column within that band) and hashes them to the buckets using a hash function. The more similar two columns are, the more likely they collide into some bands. When two items fall into the same bucket, it means a portion of their signature agrees, and they will be added to the candidate set. The candidate set returned by an LSH query only contains a subset of items that are more likely similar (e.g., having Jaccard similarity over a certain threshold). An approximation of this threshold is defined as $ST = (1/b)^{(1/r)}$. 

Typically, LSH is configured with a high $ST$ so that the candidate set only contains closely similar items. However, in our context, we are interested in items with a maximum distance from the LSH query. Thus, like Miranda et al. \cite{miranda2018fast}, we configured LSH so that we achieved an approximately 0.1 similarity threshold \footnote{permutations: 10; bands: 10; rows: 1}, and the candidate set $CS$ would contain almost all test cases, and the distant set $DS$ would include a small number of remaining items with high Jaccard distance. To employ LSH for TCP purpose, we implemented an algorithm (algorithm \ref{label_lsh_algorithm}) proposed by Miranda et al. \cite{miranda2018fast}.

\begin{algorithm}[h]
\label{label_lsh_algorithm}
\SetAlgoLined
\KwData{Test Suite $TS$}
\KwResult{Prioritized Suite $PS$}
$signatures \leftarrow$ MinHashSignature($TS$)\;
LSH.Index ($signatures$)\;
$query \leftarrow$ MinHashSignature($\emptyset$)\;
 \While{$signatures$ is not empty}{
  $CS\leftarrow$ LSH.Search($query$)\;
  $DS\leftarrow$ $signatures - CS - PS$\;
  Find $i\in DS$ with the maximum JD (estimate) to $PS$\;
  Append $i$ to $PS$ and remove from $signatures$\;
  $query \leftarrow$ Update cumulative MinHash signature of $PS$\;
 }
 \caption{Similarity-based TCP Using Locality-Sensitive Hashing}
\end{algorithm}

\subsection{Evaluation}
\label{label_evaluation}
To compare the investigated TCP techniques, effectiveness and performance are both important. Performance was measured using average method execution time (AMET) in seconds. AMET includes both the preparation time (i.e., calculating the distance matrix or LSH initialization) and the prioritization algorithm itself. To assess effectiveness, we used an APFD metric that was originally introduced by Rothermel et al. \cite{rothermel2001prioritizing} and is widely used in the literature \cite{khatibsyarbini2017test}. Let $T$ be an ordered test suite containing $n$ test cases and $F$ be a set of $m$ faults detected by $T$; then $TF_{i}$ indicates the number of test cases executed in $T$ before capturing fault $i$. APFD indicates the average percentage of faults detected and is defined as follows:

$$ APFD = 100 * (1- \frac{ TF_1 + TF_2 + ... + TF_M } { nm } + \frac{1} { 2_n }) $$
 
To properly compare the investigated TCP techniques, we performed statistical analyses. A Mann\textendash Whitney U test \cite{arcuri2011practical}, which is a non-parametric significance test, was applied to determine whether the difference between two techniques is statistically significant, using $p<0.05$ as the significance threshold. The null hypothesis of this test indicates that there is no significant difference between the effectiveness of the techniques under evaluation. This test was selected because the studied data may not follow a normal distribution. The Mann\textendash Whitney U test indicates whether there is any difference between techniques but does not show the degree of difference between them. Thus, we used a VDA measure \cite{arcuri2011practical}, which is a non-parametric effect size. A VDA measure is a number between 0 and 1. When $VDA(x, y)=0.5$, it indicates the two techniques are equal. When $VDA(x, y)>0.5$, it means $x$ outperformed $y$ and vice versa. To compare the investigated techniques across the subject programs, we presented the mean for the analyzed variables, and a 95\% non-parametric confidence interval (CI) based on 1000 bias-corrected and accelerated bootstrap replicates. Furthermore, when comparing TCP techniques, we also provided violin plots to visualize the distribution of APFDs.

\section{Findings}
This section is structured to address the research questions and includes the aggregated results of all execution rounds (see the number of versions presented for each subject in Table \ref{table:subjects}). The experiments were conducted on a computer with an Intel 2.7 GHz Xeon E5-2680 CPU and 16 GB installed RAM. To accelerate the performance of investigated TCP techniques, we parallelized all techniques.

\subsection{RQ1: Is prioritization by similarity-based TCP more effective at finding defects than random permutation?}

\begin{figure}[h]
\centering
\includegraphics[width=\textwidth]{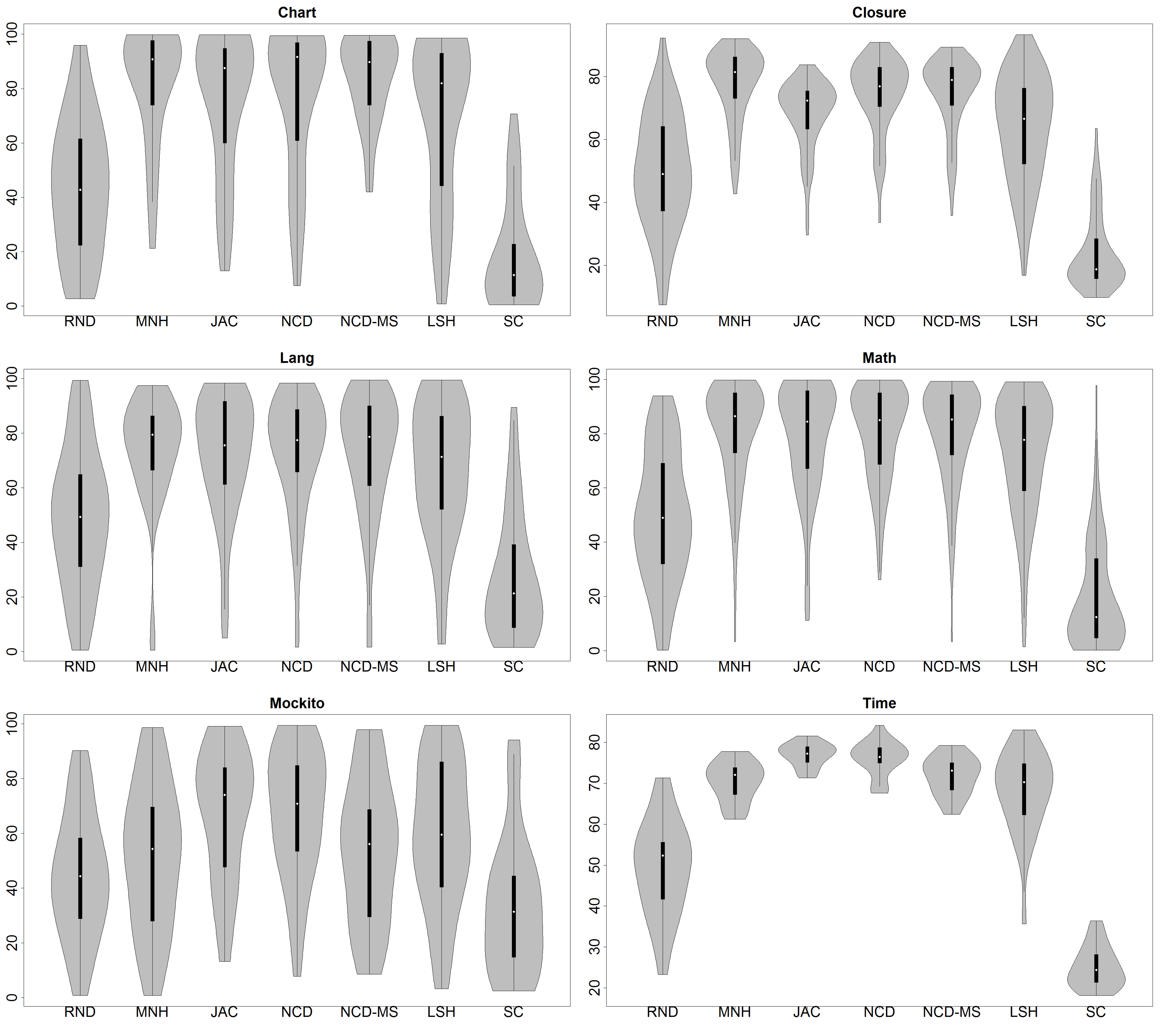}
\caption{Effectiveness (APFD) Comparison - violin plots} \label{boxplot}
\end{figure}

Table \ref{table:RQ1} presents the effect sizes for differences between analyzed SBTP techniques and random permutation. The analyzed SBTP techniques’ effectiveness varies among subjects. However, one can observe that SBTP is largely more effective in finding defects than random permutation (VDA 0.76\textendash0.99 observed using NCD across all subjects). These differences are also statistically significant in nearly all cases, which indicates running the most dissimilar test cases early in the testing process (maximizing the diversity) increases the test suite's fault-detection capability. This was also verified by our sanity check (SC) where the inverse approach was employed. The sanity check indicated maximizing similarities among tests would decrease the test suite's fault-detection capability, and as expected, it was less effective than random ordering (VDA: 0.03\textendash0.34). Figure \ref{boxplot} shows the violin plots for the investigated TCP techniques within the studied subjects.

\begin{table}[h]
\centering
\caption {VDA Effect Size - TCP Technique vs. RND Permutation}
\label{table:RQ1}
\begin{tabular}{|l|l|l|l|l|l|l|}
\hline                       
 Project & MNH & JAC & NCD & NCD-MS & LSH & SC \\
\hline
\hline 
Chart & 0.87 \cellcolor{Gray}& 0.81 \cellcolor{Gray}& 0.81 \cellcolor{Gray}& 0.91 \cellcolor{Gray}& 0.76 \cellcolor{Gray}& 0.21 \cellcolor{Gray} \\ 
Closure & 0.89 \cellcolor{Gray}& 0.8 \cellcolor{Gray}& 0.87 \cellcolor{Gray}& 0.87 \cellcolor{Gray}& 0.71 \cellcolor{Gray}& 0.12 \cellcolor{Gray} \\
Lang & 0.79 \cellcolor{Gray}& 0.75 \cellcolor{Gray}& 0.77 \cellcolor{Gray}& 0.76 \cellcolor{Gray}& 0.69 \cellcolor{Gray}& 0.25 \cellcolor{Gray} \\
Math & 0.84 \cellcolor{Gray}& 0.82 \cellcolor{Gray}& 0.84 \cellcolor{Gray}& 0.84 \cellcolor{Gray}& 0.76 \cellcolor{Gray}& 0.16 \cellcolor{Gray} \\
Mockito & 0.58 & 0.74 \cellcolor{Gray}& 0.76 \cellcolor{Gray}& 0.6 & 0.67 \cellcolor{Gray}& 0.34 \cellcolor{Gray} \\
Time & 0.96 \cellcolor{Gray}& 0.99 \cellcolor{Gray}& 0.99 \cellcolor{Gray}& 0.96 \cellcolor{Gray}& 0.9 \cellcolor{Gray}& 0.03 \cellcolor{Gray} \\
\hline
VDA Range & 0.58\textendash0.96 & 0.74\textendash0.99 & 0.76\textendash0.99 & 0.60\textendash0.96 & 0.67\textendash0.90 & 0.03\textendash0.34\\
\hline
\end{tabular}
\end{table}

\subsection{RQ2: Which similarity-based TCP technique is the most effective and has the best performance?}

For a TCP approach to be applicable in a real-world environment, effectiveness (measured by APFD) and performance (measured by AMET) are both critical. Table \ref{table:RQ2_apfd} compares the effectiveness of the investigated techniques within and across the studied subjects. However, Table \ref{table:RQ2_amet} compares the investigated techniques' performance within and across the studied projects. One can observe that all SBTP techniques except LSH achieved very close mean APFD scores across all subjects (72.69 \textendash 75.44). LSH achieved the lowest effectiveness (66.79), but had the best performance and scored a very low AMET across all subjects (1.24 seconds). Overall, on average and across all subjects, no technique was found to be superior with respect to the effectiveness. LSH was, to a small extent, less effective than other SBTP techniques (VDA: 0.38 observed in comparison to NCD), but its speed largely outperformed the other techniques (i.e., it was approximately 5 \textendash 111 times faster).

\begin{table}[h]
\centering
\caption {Effectiveness (APFD) Comparison}
\label{table:RQ2_apfd}
\begin{tabular}{|l|l|l|l|l|l|l|}
\hline
 Project & MNH & JAC & NCD & NCD-MS & LSH \\
\hline
\hline
Chart & 81.89 & 75.72 & 77.3 & 84.32 & 69.94 \\
Closure & 77.47 & 68.24 & 74.41 & 74.84 & 63.65  \\
Lang & 74.26 & 72.14 & 73.16 & 72.69 & 66.85  \\
Math & 80.39 & 78.81 & 80.37 & 80.32 & 72.21  \\
Mockito & 51.69 & 67.09 & 67.58 & 53.82 & 59.52  \\
Time & 70.39 & 76.95 & 75.99 & 71.8 & 68.06  \\
\hline
\specialcell{Mean APFD \\(95\% CI)} & \specialcell{75.05 \\(72.90\textendash76.87)} & \specialcell{72.69 \\(70.84\textendash74.47)} & \specialcell{75.44 \\(73.83\textendash77.28)} & \specialcell{74.35 \\(72.51\textendash76.26)} & \specialcell{66.79 \\(64.64\textendash68.81)}  \\
\hline
\end{tabular}
\end{table}

\begin{table}[h]
\centering
\caption {Performance (AMET) Comparison}
\label{table:RQ2_amet}
\begin{tabular}{|l|l|l|l|l|l|l|}
\hline
 Project & MNH & JAC & NCD & NCD-MS & LSH \\
\hline
\hline
Chart & 138.22 & 25.02 & 15.65 & 102.99 & 2.84  \\
Closure & 230.15 & 15.05 & 5.85 & 89.99 & 1.32  \\
Lang & 33.18 & 2.76 & 0.58 & 7.22 & 0.29 \\
Math & 142.64 & 17.22 & 10.01 & 97.69 & 1.62  \\
Mockito & 41.54 & 7.14 & 5.63 & 14.32 & 1.04  \\
Time & 56.78 & 5.88 & 1.27 & 18.24 & 0.39  \\
\hline
\specialcell{Mean AMET \\(95\% CI)} & \specialcell{138.21 \\(130.50\textendash146)} & \specialcell{12.88 \\(12.14\textendash13.57)} & \specialcell{6.41 \\(5.87\textendash6.95)} & \specialcell{67.11 \\(61.85\textendash73.29)} & \specialcell{1.24 \\(1.15\textendash1.33)} \\
\hline
\end{tabular}
\end{table}

\section{Discussion}

\subsection{Overview of Findings, Their Implications, and Future Works}

The ultimate objective of our study was to detect regression faults early in the testing process, allowing software developers to perform regression testing more frequently and continuously. To achieve our objective, we conducted an experiment using the defects4j dataset \cite{just2014defects4j}. 

Test suites ordered by SBTP were largely more effective at finding defects than random permutation (VDA: 0.76\textendash 0.99 observed using NCD across all subjects). This indicates running the most dissimilar test cases early in the testing process (maximizing the diversity) increases the test suite's fault-detection capability. This is also verified by our sanity check where the reverse approach was applied (VDA: 0.03\textendash 0.34). Of the 5 SBTP implementations investigated, no technique was found to be superior with respect to the effectiveness. LSH was, to a small extent, less effective than other SBTP techniques (VDA: 0.38 observed in comparison to NCD), but its speed largely outperformed the other techniques (i.e., it was approximately 5\textendash111 times faster). From practical perspective, NCD seems to be the best choice because it achieved high effectiveness with relatively low average method execution time. Yet, LSH is more practical when the prioritization time is critical.

Findings from our study bring to mind the well-known adage \enquote{don't put all your eggs in one basket}. To effectively consume a limited testing budget, one should spread it evenly across different parts of the system by running the most dissimilar test cases early in the testing process. The underlying intuition is that \textit{test cases that capture the same faults tend to be more similar to each other, and test cases that capture different faults tend to be more different} \cite{jiang2009adaptive,hemmati2010reducing,ledru2012prioritizing}. In comparison to other TCP techniques, SBTP requires minimal information (i.e., only the required information is encoded in the test suite) and has potential applications. SBTP can be applied in different contexts and during initial testing where no information about the system under test is available (e.g., code coverage or historical data). SBTP is an especially interesting approach when code instrumentation is too costly or impossible (e.g., in automotive system testing where source-code is not always available \cite{haghighatkhah2017improving,haghighatkhah2017automotive}). SBPT can also be applied in a complementary fashion and combined with other TCP techniques (e.g., history-based diversity proposed in our previous work \cite{haghighatkhah2018test}).

To realize SBTP in practice, one must measure the similarities among test cases. This similarity measurement can be performed using string metrics and on different properties (i.e., the source code, documentation, or any other information about the test cases). As acknowledged by Ledru et al. \cite{ledru2012prioritizing}, string metrics are based on lexicographic information and do not necessarily capture the semantics behind the test cases. Two test cases might consequently be considered similar, although they are distant and correspond to different execution paths. Future works are required to investigate possible approaches that precisely measure the semantic similarities among test cases. The candidate approach should not come with a high overhead; otherwise, its application remains in theory.

Once similarity measurement has been performed, this information should be leveraged to perform TCP. One can argue that diversification is perhaps the best strategy when no strong clues about fault-revealing test cases are available. Test diversity is a classical heuristic in the literature and has been applied previously \cite{jiang2009adaptive,hemmati2010reducing,ledru2012prioritizing,hemmati2013achieving,hemmati2017prioritizing,arafeen2013test,thomas2014static,feldt2016test,flemstrom2017similarity}. The opposite viewpoint is the intensification strategy, where the testing budget is consumed by and around the most probable fault-revealing test cases. Theoretically, both strategies can be applied simultaneously (i.e., intensify where it is necessary and diversify the remaining budget). However, making decisions about when and how to apply these strategies, either individually or combined, remains a challenge. To the best of our knowledge, the application of these strategies, as well as their relevance and impact, have not been widely investigated in the literature. The only exception we are aware of is the recent study by Patrick and Jia \cite{patrick2017kd} wherein the authors investigated the trade-off between diversification and intensification in adaptive random testing.

Regardless of which strategy is chosen, a TCP algorithm needs to iteratively find the most (dis)-similar item to the set of already prioritized test cases. This can be done using different search techniques. TCP using a pairwise algorithm does not scale, and its performance becomes inefficient as the test suite's size increases. In this work, we have investigated LSH as one popular solution to the similarity search problem. There are other solutions proposed in the literature. Future work should also investigate the effectiveness and performance of candidate solutions.

\subsection{Threats to Validity}
In empirical software engineering, validity threats can be grouped into four distinct classes: construct validity, internal validity, external validity, and reliability \cite{wohlin2000experimentation}. In the present context, construct validity relates to the use of right measures. To assess the investigated TCP techniques' effectiveness, we used the APFD metric, which is widely used in the literature (see the latest systematic literature review on TCP by Khatibsyarbini et al. \cite{khatibsyarbini2017test}). Internal validity concerns the relationship between the constructs and the proposed explanation. This corresponds to the potential faults in our implementation. Our implementation was piloted on a small sample before running the actual experiment. Furthermore, the implementation and results were discussed and reviewed in regular meetings, which were held among the co-authors of this study. 

External validity relates to the generalizability of the study and whether the subjects of our study are real-world projects. Our experiment was conducted on the defects4j dataset \cite{just2014defects4j}, which contains 395 real faults from 6 real-world open-source Java programs. Our conclusions are drawn based on ex-post analysis of software artifacts. This motivates our future work to replicate our experiment in industry and to larger systems. Reliability concerns the repeatability and reproducibility of the research procedure and conclusions. This required access to the analyzed subjects and a throughout report of the experiment. The data that we used is publicly available, and detailed information about our experiment and its implementation were presented in this paper.

\section{Concluding Remarks}

The ultimate objective of our study was to detect regression faults early in the testing process, allowing software developers to perform regression testing more frequently and continuously. To achieve this objective, we conducted an experiment using the defects4j dataset, which contains 395 real faults from 6 real-world open-source Java programs. In summary, the results from our experiments suggest the following: 

(1) Test suites ordered by SBTP were largely more effective at finding defects than random permutation (VDA: 0.76\textendash 0.99 observed using NCD across all subjects), which means running the most dissimilar test cases early in the testing process improves the test suite's fault-detection capability; (2) Of the 5 SBTP implementations investigated, no technique was found to be superior with respect to the effectiveness. LSH was, to a small extent, less effective than other SBTP techniques (VDA: 0.38 observed in comparison to NCD), but its speed was faster than the other techniques studied (approximately 5\textendash 111 times faster).

Taken together, these results bring to mind the well-known adage \enquote{don't put all your eggs in one basket}. To effectively consume a limited testing budget, one should spread it evenly across different parts of the system by running the most dissimilar test cases early in the process. Our study contributes to the literature by providing empirical evidence in support of test diversity and its impact on TCP.

%
%
%

\bibliographystyle{splncs04}
\bibliography{profes18}

\end{document}